\documentclass{appolb}
\usepackage{epsfig}

\newcommand{\beq}{\begin{equation}} 
\newcommand{\eeq}{\end{equation}}   
\newcommand{\bea}{\begin{eqnarray}} 
\newcommand{\eea}{\end{eqnarray}}

\begin{document}
\date{\today}
\pagestyle{plain}
\newcount\eLiNe\eLiNe=\inputlineno\advance\eLiNe by -1
\title{ Radiative return physics program within EURIDICE network
\thanks{Presented by H.~Czy\.z 
at The Final EURIDICE Meeting "Effective theories of colours
  and flavours: from EURODAPHNE to EURIDICE", Kazimierz, Poland,
  24-27 August, 2006.
 Work supported in part by EC 5th Framework Program under contract
HPRN-CT-2002-00311 (EURIDICE network),
 EC 6-th Framework Program under contract
  MRTN-CT-2006-035482 (FLAVIAnet),  TARI project RII3-CT-2004-506078
 and
  Polish State Committee for Scientific Research
  (KBN) under contract 1 P03B 003 28.}
}
\author{Henryk Czy\.z$^a$ and Agnieszka Grzeli\'nska$^b$
\address{ a: \ Institute of Physics, University of Silesia,
PL-40007 Katowice, Poland \\ b: \ Institut f\"ur Theoretische Teilchenphysik,
Universit\"at Karlsruhe,\\ D-76128 Karlsruhe, Germany
 }}
\maketitle

\begin{abstract}
 A short review of both theoretical and experimental aspects 
 of the radiative return method is presented with the emphasize
 on the results obtained within the EURIDICE network. It is shown
 that the method gives not only possibility of an independent,
 from the scan method, measurement of the hadronic cross section,
 but also can provide information concerning details of the hadronic
 interactions.  
 \end{abstract}
\PACS{13.40.Ks,13.66.Bc}

\section{Introduction}
  
 Four years of a very active physics program of the EURIDICE network was
 also very fruitful for the group developing the radiative return method
 \cite{Zerwas}
 and tools necessary for the physical analysis. Major part of all theoretical
 investigations in that topic was done within this network, thus the review
 of the theoretical investigations is in fact the review of the results 
  obtained within the EURIDICE network. Software developed and/or
  upgraded within the network:
  PHOKHARA\cite{Szopa,rest,Czyz:PH03,Nowak,Czyz:2004nq,PHOKHARA:mu}
  and EKHARA \cite{Czyz:2006dm}
  generators together with the event generators
  developed by the same group (EVA \cite{Binner}, EVA4pi \cite{CK}) 
   prior to the starting date of
  the EURIDICE is successfully used by experimental groups working at meson
  factories BaBar and KLOE and soon will be used also at BELLE.

   The method originally developed for the hadronic cross section
 measurement was  successfully used by KLOE \cite{KLOE2}
 to obtain $\sigma(e^+e^-\to\pi^+\pi^-$) with the precision competitive
 to the one obtained by means of the scan method. It was also used by BaBar to
 obtain cross sections of many hadronic channels which were measured
 with poor accuracy or were not measured previously. Let us mention only 
 few of them: narrow resonances studies \cite{Aubert:2003sv}, 
 big improvement in the accuracy of the three pion \cite{Aubert:2004kj}
 and four charged mesons (pions and kaons) \cite{Aubert:2005eg} cross sections
 measurements and proton form factors extraction \cite{Aubert:2005cb}.
 
 The paper starts with a description of the radiative return method in 
 Section~\ref{sec2} and the ingredients necessary for the precision
  physics in Section~\ref{sec3}. The final state photon emission (FSR), is
 discussed in Section \ref{sec4}, together with methods how to handle it
 and to be able to perform precision measurements.
 Selected topics beyond the hadronic cross section measurements 
 are described in Section \ref{sec5}.

\section{The basics of the radiative return method \label{sec2}}

 Simple and innovative observation made some years ago \cite{Zerwas}
 lead, with series of papers which started with \cite{Binner} and \cite{CK},
 to a development of a radiative return method giving
 today many valuable physical results. The method gives an access to 
 information contained in the hadronic cross section 
 $d\sigma(e^+e^-\to{\rm hadrons})$ through a measurement of the hadronic
 invariant  mass distribution in the reaction
  $e^+e^-\to{\rm hadrons}+{\rm photons}$.
  Historically the process of $e^+e^-$ annihilation to a pair 
(or arbitrary number) of particles
 plus one photon was investigated earlier 
\cite{Baier:1965jz,Baier:1965bg,Fadin}, but the scope of 
 that papers was not to provide with a method to measure the hadronic 
 cross section.

\begin{figure}[ht]
\begin{center}
\epsfig{file=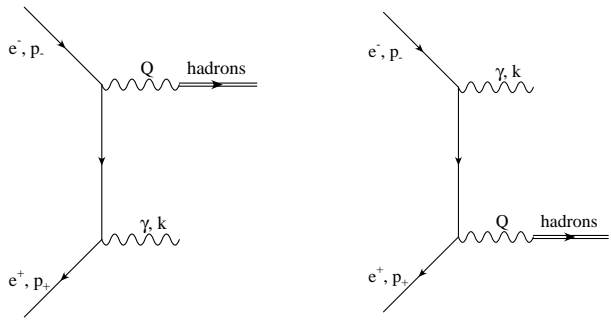,width=8.cm} 
 \\
\caption{The leading order diagrams contributing to the radiative return
cross section. }
\label{LO}
\end{center}
\end{figure}

To illustrate in detail how the method works let us consider the lowest order
 contribution to the radiative return cross section. The  contributing
diagrams are shown in Fig. {\ref{LO}}, where only initial state radiation
 (ISR) is taken into account. The complications caused by final state radiation
 (FSR), as well as the methods to overcome them, are discussed in 
 Section \ref{sec4}.

The corresponding ISR matrix element has the following form

\begin{eqnarray}
 {\cal M} \sim  \bar v\left(p_{+}\right) 
 \left[ \gamma^{\nu} 
 \frac{1}
   {p_{-}{\kern-12pt}/ \ - k {\kern-5pt}/ - m}
     \epsilon^{*}{\kern-9pt}/\ (k) 
 + \epsilon^{*}{\kern-9pt}/ \ (k) 
    \frac{1}{k {\kern-5pt}/ - p_{+}{\kern-12pt}/ \ - m}\gamma^{\nu}
 \right] u(p_-) \  
 \frac{1}{Q^2}\ J^{em}_{\nu} \ ,
\label{matrix}
\end{eqnarray}

\noindent
where  $J^{em}_{\nu}$ is the electromagnetic hadronic current present
also in the matrix element describing  the process $e^+e^-\to{\rm hadrons}$. 
 From Eq.(\ref{matrix}) it is clear that a factorization, allowing separation
 of the hadronic part, will take place even for the
 squared matrix element. Indeed, integrating the cross section 
 over the hadrons phase space $d\bar\Phi_n$ ('bar' indicates that all
 statistical factors are included in its definition) one gets 
\begin{eqnarray}
 \int \ { J^{em}_\mu (J^{em}_\nu)^*}  \ \ d\bar\Phi_n(Q;q_1,\dots,q_n) = 
 \frac{1}{6\pi} { \left(Q_{\mu}Q_{\nu}-g_{\mu\nu}Q^2\right)}
  \ { R(Q^2)} \ ,
 \label{hadronict}
\end{eqnarray}
\noindent
where
\begin{eqnarray}
R(s)=\frac{\sigma(e^+e^-\rightarrow hadrons)}{\sigma_{\rm point}}
 \ , \ \sigma_{\rm point} = \frac{4\pi\alpha^2}{3s} \ .
\label{rratio}
\end{eqnarray}

That leads to the relation between cross sections with and without
 photon emission

 \begin{eqnarray}
 d\sigma(e^+e^- \to \mathrm{hadrons} + \gamma ({\rm ISR})) &=&
\nonumber \\ 
  &&\kern-50pt
 H(Q^2,\theta_\gamma) \  d\sigma(e^+e^-\to \mathrm{hadrons}, s=Q^2) \ ,
\label{factorization}
\end{eqnarray}
\noindent
where $Q^2$ is the invariant mass of the hadronic system.
As it is clear from Eq. (\ref{matrix}) and Eq. (\ref{hadronict})
the factorization of the Eq. (\ref{factorization}) remains valid
 at any order provided that only ISR is considered. The function
 $H(Q^2,\theta_\gamma)$ at relatively low energies of meson factories
 is given with high accuracy by QED only and thus it is well
 known. That means that by measurement of the
 differential in $Q^2$ cross section of the reaction
 $e^+e^- \to \mathrm {hadrons} + \mathrm{photons}$
 one can extract the cross
 section $e^+e^- \to \mathrm{hadrons}$  for the energies from the 
 production threshold to almost the nominal energy of a given experiment,
 provided that one is able to overcome complications
 described in Section~\ref{sec4}. The cross section
 of the reaction with photon emission is lower then the one without
 photon emission, thus its measurements require higher luminosity then the ones
 necessary for scan experiments for similar statistical accuracy.
 However, it does not require to build a dedicated experiments
 and it can use data collected at any of the meson factories, where
 luminosity is not a problem.

\section{Ingredients necessary for a precise analysis \label{sec3}}

  An extraction of the hadronic cross section via the radiative return
 requires a few basic ingredients:
 \begin{itemize}
  \item{ an accurate calculation of the ISR 
  including appropriate radiative corrections}
  \item{an adequate, tested experimentally, model of the FSR}
   \item{a Monte Carlo event generator to be able to use the theoretical
  information in a realistic experimental set up
  } 
   \item{$e^+e^-$ scattering experiment with a high luminosity
 and a good detector}
 \end{itemize}

 The virtual radiative corrections to ISR were calculated in \cite{PHradcor}
while the real emission was included in the developed Monte Carlo generator
PHOKHARA in \cite{Szopa} and \cite{rest}. The estimated 
physical accuracy of the ISR contributions, 0.5\%, was adequate at the
 time of the release of the generator even for the precise KLOE 
 pion form factor extraction \cite{KLOE2}.
 The new data collected at DAPHNE by KLOE collaboration 
\cite{Ambrosino:2007nx,Leone:2006bm},
 together with an improvement in the theoretical description of the
 Monte Carlo event generator BABAYAGA \cite{Balossini:2006wc}, requires however
 further work and an inclusion of the NNLO contributions.
 From the estimates that were performed, taking into account the 
 leading logarithmic corrections coming from the second loop 
  together with one loop leading logarithmic
  corrections to the cross section with two photon emission
  and three ISR photon emission,
  will be enough to reach the precision of 0.1-0.2\%.
  That is a necessary condition to be able
 to fully profit from the new very accurate data.  

  The FSR modeling requires close collaboration between experimental
 and theory groups to reach required precision in a short time and
 will be discussed in the next Section.

 A development of a reliable Monte Carlo generator requires not only
 the precise knowledge of both ISR and FSR, but a continuous work on
 efficiency of the generator and its tests for each added hadronic
 channel. Only this way one can assure reliability of the developed
 product used afterwords by demanding experimental groups.

\section{Final state photon emission: the problems and how to overcome them
  \label{sec4}}

 The final state emission (FSR) forms a potential problem for the application
 of the radiative return method and it has to be studied carefully
 to assure adequate accuracy of the description.
 At 
 B- factories the  region of
 hadronic masses of physical interests,
 below 4~GeV, lays far from the nominal energy of the
 experiments and an emission of a hard photon is required to
  reach it. It means that the typical kinematic configuration of an 
 event consists of a photon emitted in one direction and hadrons
 going opposite to it. That suppresses
 the FSR contributions, which are large for photons emitted parallel
 to the direction of a charged hadron in the final state, and makes
 the measurement of the hadronic cross section easier. For the $\phi$- factory
 DAPHNE,
 where the region of interest is not far from the nominal
 energy of the experiment, that natural separation between the emitted
  photon and the hadrons does not take place and one has to suppress FSR
 by an appropriate event selection. As a result one has to control
 the uncertainty due to the model dependence of the final state emission.
 That forms a challenge, as the existing models were not tested
 with the adequate
 precision prior to the DAPHNE results. Let's discuss that problem
 on the basis of the $e^+e^-\to \pi^+\pi^-\gamma(\gamma)$ reaction,
 where the accuracy is the most demanding. 
 The solution  was first proposed in \cite{Binner}
 and further elaborated in \cite{Czyz:PH03}.
 A similar
 investigations is possible for other hadronic final states,
 however till now it
 was not performed. 
 
 The main tool in the tests of the model(s) of the photon
 emission from the final pions is the charge asymmetry.
 For ISR emission of any
 number of photons the two-pion state is produced in C=-1 state and
 with an 
 odd orbital angular momentum.
 For one real photon emitted from the final pions
  the two-pion state is produced
in C=1 state and with an even orbital angular momentum. 
  As a result, the initial-final
 state interference for one photon emission 
 is odd under $\pi^+\leftrightarrow \pi^-$ interchange
 and the integrals for charge blind event selections are equal to zero.
 In the same time the interference
 is the only source of the charge asymmetry and  allows for
 tests of the models of the final state emission. The charge asymmetry
  depends on the invariant mass of the two-pion system and that
 allows for  detailed tests of the model(s) of the FSR emission. 
 In short, the tests should be done in the following way: First
 one compares the experimental data  for the asymmetry with the Monte Carlo
 where the tested model was implemented. That has to be performed 
 for an event selection which enhance the FSR as compared to the ISR.
 Once the implemented model agrees with the data one chooses an 
 event selection, which suppresses the FSR and one measures the 
 radiative cross section for that event selection. The described procedure
 guaranties that the
 ISR and the FSR contributions are separately well under control.
 For the case of untagged photons a specific background,
 $e^+e^-\to\pi^+\pi^-e^+e^-$, has to be also taken into account
 \cite{Hoefer:2001mx,Czyz:2006dm} as the final leptons are not vetoed and
 even if they are, the major part of them escapes detection being emitted
 at small angles.
 
 The  reaction
 $e^+e^-\to \pi^+\pi^-\gamma$,
  with the photon emitted from the pions, does contribute also to 
 dispersion integrals for evaluation of $a_\mu$ and $\alpha_{QED}$.
 In the former case its theoretically estimated value \cite{Czyz:PH03}
 is about 1.5 times higher then
  the size of the present 
  theoretical uncertainty \cite{Jegerlehner:2007xe,eidelman2007}
  and thus numerically
 important. The sketched program was successfully undertaken
 by KLOE and resulted in a sound extraction of the 
 $\sigma(e^+e^-\to \pi^+\pi^-)$ \cite{KLOE2} together
 with the mentioned photon corrections. The most general form
 of the photon emission from scalar particles produced in $e^+e^-$ 
annihilation, with three form factors was investigated already
 in \cite{Baier:1965jz}, where it was shown also that only one
 form factor is relevant in the limit of soft photon emission.
 Physical program initiated there was undertaken in 
\cite{Pancheri:2006cp,Pancheri:2007xt}
 with a special emphasize on the FSR tests at KLOE.

  Another source of complications for using of
 the radiative return method
 at DAPHNE are the radiative $\phi$ decays. That problem
 was considered for the first
 time in \cite{Melnikov:2000gs} and it is discussed in more details 
 in the next section. 

\section{Not only the hadronic cross section \label{sec5}}

\begin{figure}[ht]
\begin{center}
\epsfig{file=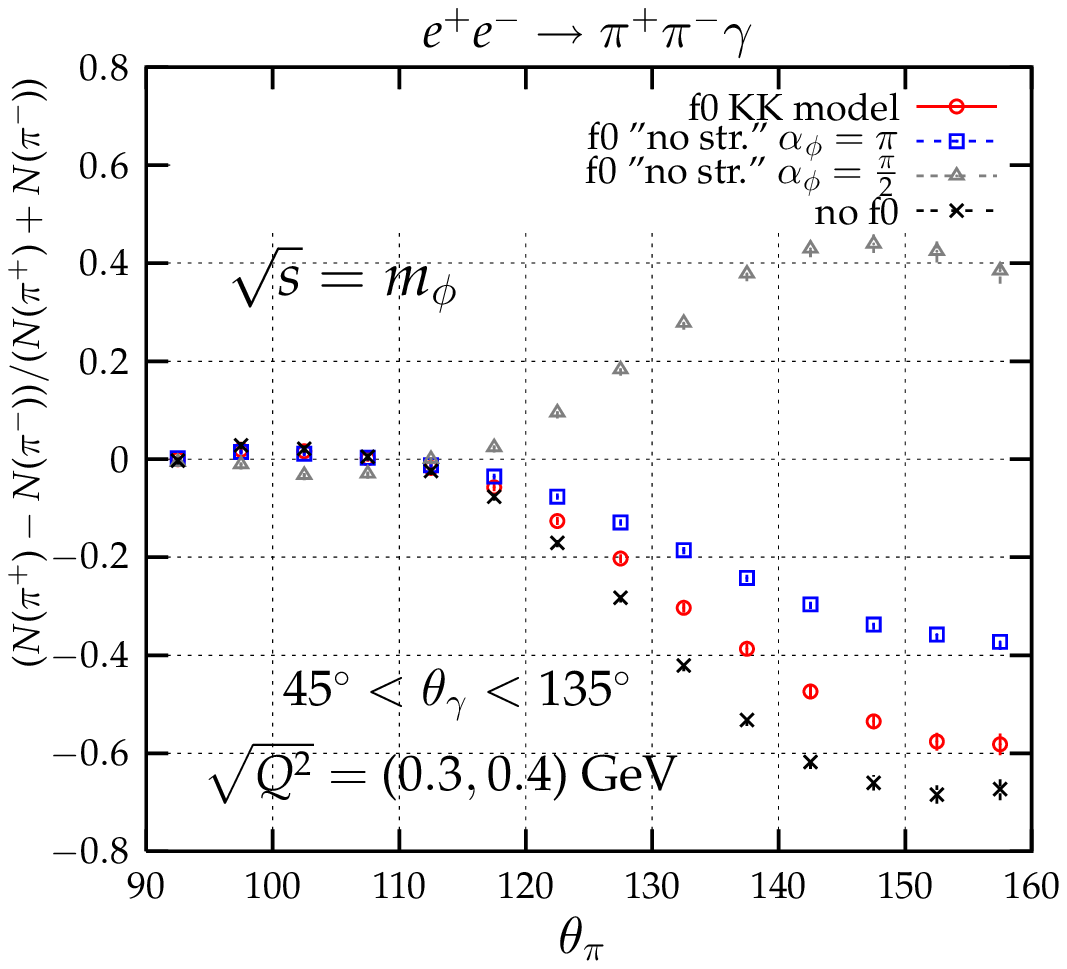,width=6cm} 
\epsfig{file=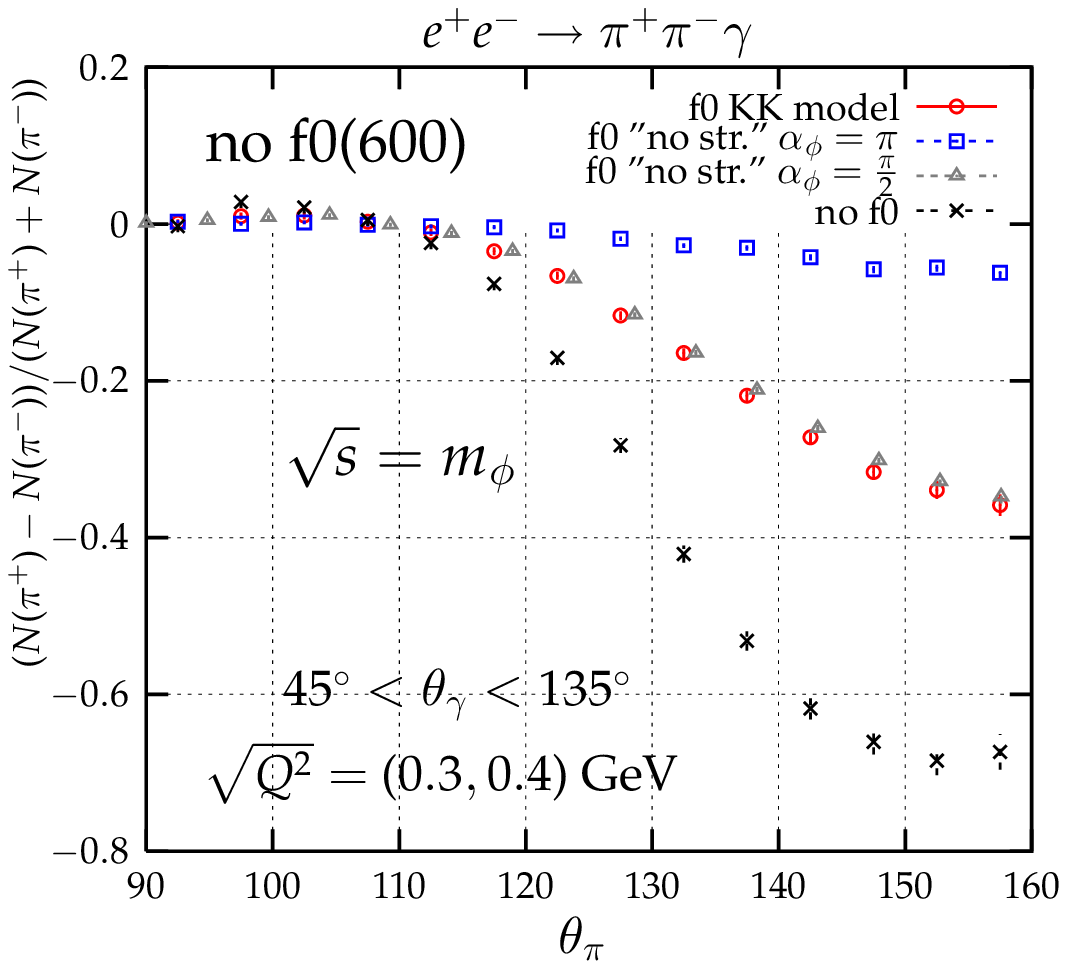,width=6cm} 
\caption{ Charge asymmetries for three different models of the radiative 
$\phi$ decays compared to the result based on sQED only (no $f_0$).
 Left plot with $f_0(980)$ and $f_0(600)$, right plot with $f_0(980)$ only.}
\label{charge_asym}
\end{center}
\end{figure}
The study of the $\phi\to\pi\pi\gamma$ decays at DAPHNE is a subject, where
the problem of FSR emission in the pion form factor extraction is interlinked
 with the possibility of using of the radiative return method to study
 hadronic models of the $\phi\to\pi\pi\gamma$ decays.
 In \cite{Czyz:2004nq} charge asymmetries were proposed to test both 
 topics. The sensitivity to the model parameters, which can be reached
 this way is definitely better then the one obtained by the fit to the $Q^2$
 spectra. An example  is shown in Fig.\ref{charge_asym}, where
 charge asymmetries predicted within a number of models of the radiative
 $\phi$ decays is shown for models with and without $f_0(600)$ contribution.
 Very distinct asymmetries promise deep insight in the details of 
 the constructed models.
The asymmetry was partially used by KLOE  \cite{Ambrosino:2005wk}
 to cross check the fit of the model parameters to the $Q^2$ spectrum
 of the pion pair invariant masses. More detailed tests are expected
 at KLOE, when both data taken at the $\phi$ resonance and below it
 will be analyzed.

\begin{figure}[ht]
\begin{center}
\epsfig{file=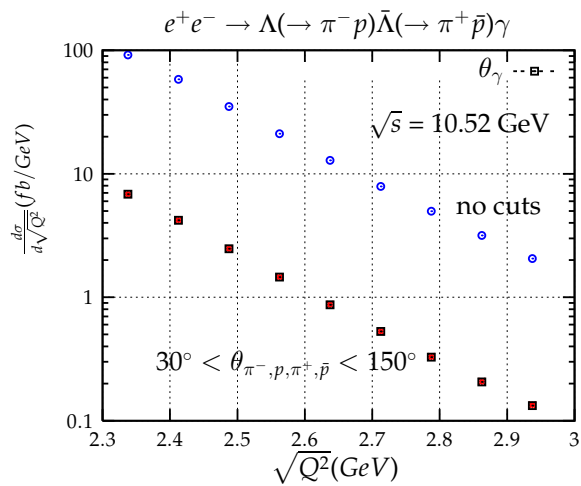,width=8.cm} 
 \\
\caption{The differential, in the invariant mass of the $\Lambda\bar \Lambda$
pairs ($Q^2$), cross section of the process  
$e^+e^- \to \Lambda (\to p\pi^-) \bar \Lambda (\to \bar p \pi^+)\gamma$
 at B-factories energy: circles - with no cuts applied, crosses - angular cuts
 on the pions and protons only, squares (overlapped with crosses)
 - angular cuts on pions, protons and the photon. }
\label{lamtot}
\end{center}
\end{figure}
Another example of using of the radiative return method to study
hadron properties is the baryon form factors extraction.
It was shown in  \cite{Nowak} that at B-factories it is possible 
 to extract the nucleon form factors up to their phases for a wide range
 of the invariant masses of the nucleon-antinucleon pairs. That is
possible through studies of the nucleon angular distributions. In a
properly chosen reference frame in which they are studied, a rest frame of the
 nucleon-antinucleon system with z-axis along the emitted photon, that
 studies are particularly simple as the angular distributions resemble
 there the nucleon angular distributions
 of the process without a photon emission. That measurement
 was successfully performed by BaBar collaboration \cite{Aubert:2005cb}.

 For the production of unstable baryon pairs,
 the decay products carry information about their
 spins and serve as spin analyzers providing with complete information
 about the production process. In \cite{Czyz:2007wi} it was studied in details
 for the process
 $e^+e^- \to \Lambda (\to p\pi^-) \bar \Lambda (\to \bar p \pi^+)\gamma$.
 The feasibility of such measurement at B-factories is obvious from 
Fig.\ref{lamtot} as one expects about one hundred events per 100 fb$^{-1}$
in the range of BaBar detector.  

\begin{figure}[ht]
\begin{center}
\epsfig{file=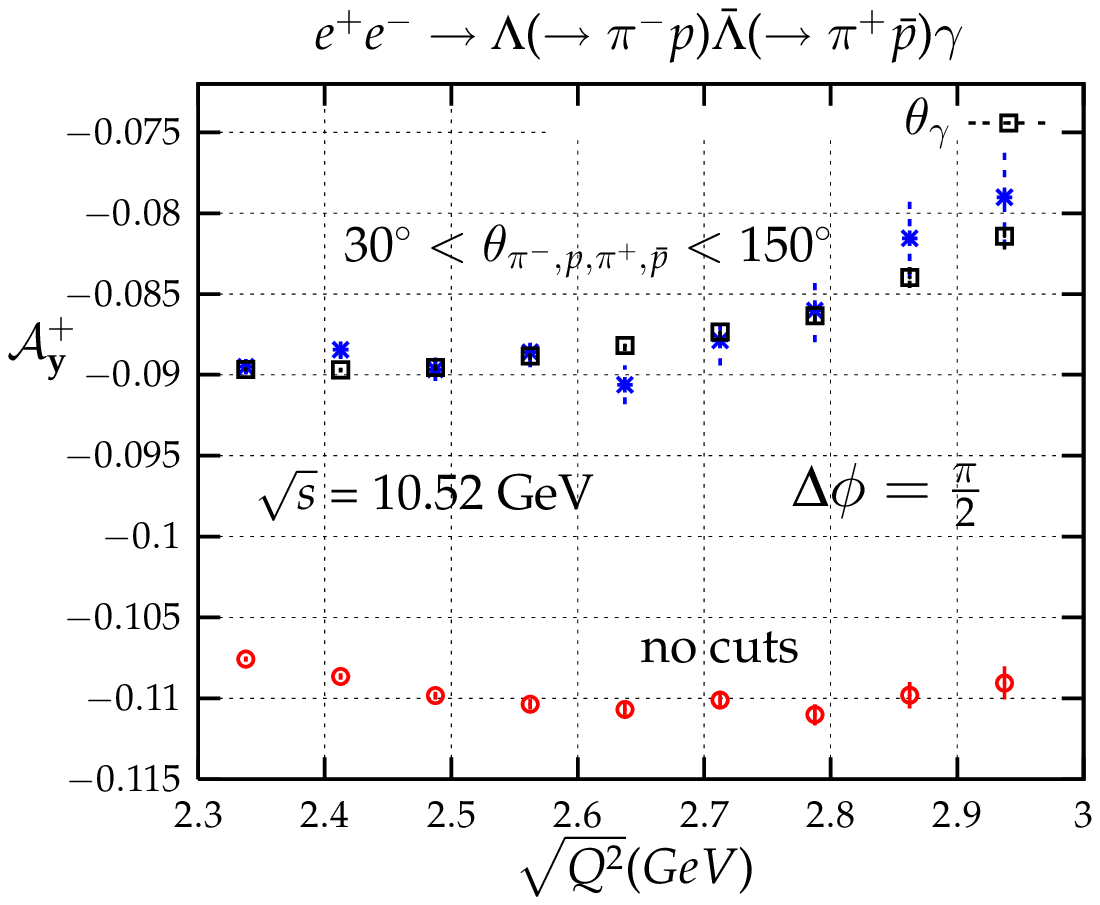,width=6.1cm} 
\epsfig{file=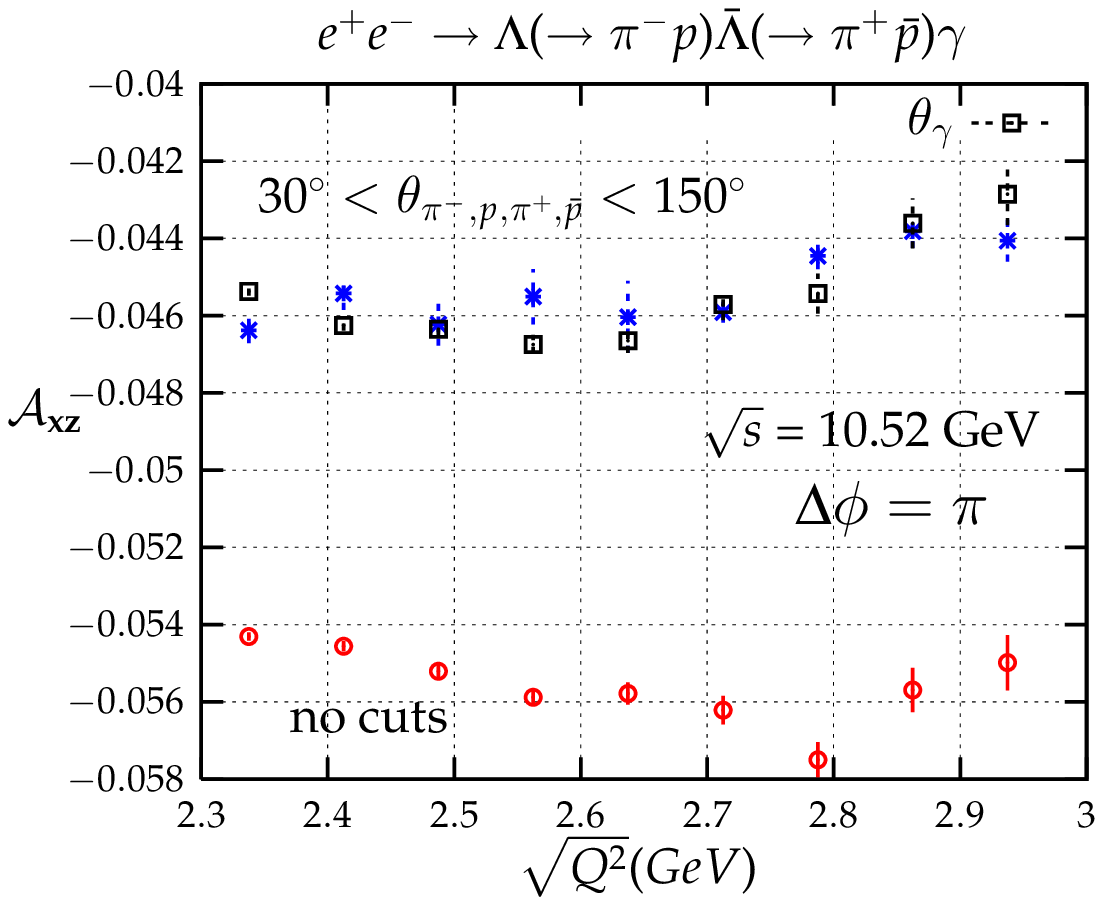,width=6.1cm} 
\caption{ Spin asymmetry (left) for the relative phase between
 magnetic and electric lambda form factors $\Delta\phi=\frac{\pi}{2}$
  and spin correlations (right) for  $\Delta\phi={\pi}$ in 
 the process 
 $e^+e^- \to \Lambda (\to p\pi^-) \bar \Lambda (\to \bar p \pi^+)\gamma$:
 circles - with no cuts applied, crosses - angular cuts
 on the pions and protons only, squares 
 - angular cuts on pions, protons and the photon.}
\label{lamspin}
\end{center}
\end{figure}

The direction of  pions coming from lambda decays is strictly correlated
with the spin of the decaying lambdas and by observing them one can measure
both spin asymmetries and spin correlations in the process of production
of the lambda-antilambda pairs.  
The spin asymmetry,
which is proportional to the sine of the phase difference of the
  magnetic and electric lambda form factors, is shown, for 
$\Delta\phi=\frac{\pi}{2}$, in Fig.\ref{lamspin} (left).
  The xz-spin correlation (please see \cite{Czyz:2007wi} for the reference
 frames definition),
which is proportional to cosine of the phase difference of the
  magnetic and electric lambda form factors, is shown, for 
$\Delta\phi= {\pi}$, in Fig.\ref{lamspin} (right).
 It is enough to measure one of them to determine
the phase difference of the lambda form factors up to a twofold sign
ambiguity and the other serves to determine that sign.
The analysis can be applied also to other members of the baryon octet. 
As almost nothing is known about their form factors,  B-factories
can provide valuable physical information allowing to test symmetries
of the underlying  models.

\section{Summary}

The radiative return research program carried within the EURIDICE network
was outlined. It was shown that just by theoretical and experimental analysis
of the data of the existing 
 meson factories one can get rich information concerning
hadronic physics, which is not limited to the hadronic cross section
 measurement. 
\vskip 0.5 cm

\centerline{***}

 The publication is based in a big part on results obtained in collaboration 
 with  
 J.~H.~K\"uhn, E.~Nowak-Kubat and G.~Rodrigo. The authors are grateful for 
many
 useful discussions concerning experimental aspects of the radiative
 return method to members of the KLOE and BaBar collaborations, mainly 
 Cesare Bini, Achim Denig, Wolfgang  Kluge, Debora Leone, Stefan M\"uller,
 Federico Nguyen, Evgeni Solodov and Graziano Venanzoni.


\end{document}